\documentclass[preprint2]{emulateapj}

\newcommand{\bzcat}{Roma-BZCAT}

\newcommand{\fer}{{\it Fermi}}

\newcommand{\wse}{{\it WISE}}

\usepackage{graphicx}
\usepackage{longtable}
\usepackage{mdwlist}
\usepackage{natbib}
\usepackage{longtable}
\usepackage{rotating}
\usepackage{morefloats}
\usepackage{threeparttable}
\usepackage{wasysym}
\usepackage{txfonts}
\usepackage{gensymb}
\usepackage{multirow}
\usepackage{upgreek}
 \usepackage{booktabs}
\usepackage[verbose]{placeins}
\usepackage{blindtext}

\begin{document}

\title{Optical archival spectra of blazar candidates of uncertain type in the 3$^{rd}$ Fermi Large Area Telescope Catalog}

\author{
N. \'Alvarez Crespo\altaffilmark{1,2},
F. Massaro\altaffilmark{1,2},
R. D'Abrusco\altaffilmark{3},
M. Landoni\altaffilmark{4},
N. Masetti\altaffilmark{5,6},   
V. Chavushyan\altaffilmark{7},                         
E. Jim\'enez-Bail\'on\altaffilmark{8},
F. La Franca\altaffilmark{9},
D. Milisavljevic\altaffilmark{3},
A. Paggi\altaffilmark{3},
V. Pati\~no-\'Alvarez\altaffilmark{7},   
F. Ricci\altaffilmark{9}
\&         
Howard A. Smith\altaffilmark{3}
} 

\altaffiltext{1}{Dipartimento di Fisica, Universit\`a degli Studi di Torino, via Pietro Giuria 1, I-10125 Torino, Italy}
\altaffiltext{2}{Istituto Nazionale di Fisica Nucleare, Sezione di Torino, I-10125 Torino, Italy}
\altaffiltext{3}{Harvard - Smithsonian Center for Astrophysics, 60 Garden Street, Cambridge, MA 02138, USA}
\altaffiltext{4}{INAF-Osservatorio Astronomico di Brera, Via Emilio Bianchi 46, I-23807 Merate, Italy}
\altaffiltext{5}{INAF - Istituto di Astrofisica Spaziale e Fisica Cosmica di Bologna, via Gobetti 101, 40129, Bologna, Italy}
\altaffiltext{6}{Departamento de Ciencias F\'isicas, Universidad Andr\'es Bello, Fern\'andez Concha 700, Las Condes, Santiago, Chile}
\altaffiltext{7}{Instituto Nacional de Astrof\'{i}sica, \'Optica y Electr\'onica, Apartado Postal 51-216, 72000 Puebla, M\'exico}
\altaffiltext{8}{Instituto de Astronom\'{\i}a, Universidad Nacional Aut\'onoma de M\'exico, Apdo. Postal 877, Ensenada, 22800 Baja California, M\'exico}
\altaffiltext{9}{Dipartimento di Matematica e Fisica, Universit\`a Roma Tre, via della Vasca Navale 84, I-00146, Roma, Italy}

\begin{abstract}
Despite the fact that blazars constitute the rarest class among active galactic nuclei (AGNs) they are the largest known population of associated $\gamma$-ray sources.
Many
of the $\gamma$-ray objects listed in the \fer-Large Area Telescope Third Source catalog (3FGL) are classified as
blazar candidates of uncertain type (BCUs), either because they show multifrequency behaviour similar to blazars
but lacking optical spectra in the literature, or because the quality of such spectra is too low to confirm
their nature.
Here we select, out of 585 BCUs in the 3FGL, 42 BCUs which we identify as probable blazars by their WISE infrared colors and which also have optical spectra that are available in the 
Sloan Digital Sky Survey (SDSS) and/or Six-Degree Field Galaxy Survey Database (6dFGS). We confirm the blazar
 nature of all of the sources. We furthermore conclude that 28 of them are BL Lacs, 8 are radio-loud quasars with flat radio spectrum and 6 are  BL Lac whose emission is dominated by their host galaxy.
\end{abstract}


%

\section{Introduction}
\label{sec:introduction}

According to the unification scenario of radio-loud active galactic nuclei (AGNs), blazars are an extreme class of extragalactic sources, whose emission is due to a relativistic jet pointed closely aligned 
towards the observer's line of sight
\cite[see e.g.][]{blandford78,urry95}. There are two main subclasses of blazars, flat spectrum radio quasars (FSRQ) and BL Lac objects. 
We label them according to the nomenclature of the \bzcat\ catalog \cite{massaro09}: if their optical spectrum is featureless or shows only optical emission/absorption lines
with equivalent widths EW $< 5 \AA$ \citep{stickel91} the blazar is classified as BL Lac (i.e., BZB), while if it shows a typical quasar-like optical spectrum it is labeled as blazar of quasar type (i.e., BZQ). 
We also indicate BL Lacs exhibiting optical spectra of
a typical elliptical galaxy (with a low Ca H\&K break contrast) as BL Lacs of galaxy type (BZGs, see e.g. Massaro et al. 2012a, 2015 for details).

Although they have a low sky density ($\sim$ 0.1 sources/degree$^2$), blazars constitute the most numerous
population of extragalactic  $\gamma$-ray sources.
According to the Third catalog of active galactic nuclei detected by the \fer- Large Area Telescope (3LAC, Ackermann et al. 2015),
29\% of the sources detected are FSRQs and 41\% BL Lacs.
Also 28\% of the sources detected above 100 MeV
present multifrequency behaviour similar to blazars (e.g. flat radio spectra and/or X-ray counterpart, see e.g. Ackermann et al. 2015 and references therein),
but often no optical spectra are available to precisely determine its class, or their signal-to-noise ratio could be too low to allow us a precise determination of their nature \cite{ackermann12}.
When such situation occur $\gamma$-ray objects are classified as blazar candidates of uncertain type (BCUs). 
 These sources belong to the $\gamma$-ray class labelled as active galaxies of uncertain type (AGUs)
in the first and second releases of the \fer-LAT Point Source Catalog (1FGL,  Abdo et al. 2010 and 2FGL, Nolan et al. 2012). 
According to the 3LAC, the BCU sources are divided into three sub-types:
\begin{itemize}
\item BCU I: The counterpart has a published optical spectrum but it is not sensitive enough for a classification as a FSRQ or a BL Lac.
\item BCU II: The counterpart is lacking an optical spectrum but a reliable evaluation of the SED synchrotron-peak position is possible.
\item BCU III: The counterpart is lacking both an optical spectrum and an estimated synchrotron-peak position but shows blazar-like broadband emission and a flat radio spectrum.
\end{itemize}

Using the infrared (IR) colors as proxy of the blazar-like behavior, Massaro et al. (2012b) showed that BCUs feature the same
IR emission as known $\gamma$-ray blazars \citep{massaro11,dabrusco12}.  Motivated by this result, a search of IR selected blazar candidates was also extended
to include the unidentified/unassociated $\gamma$-ray sources \citep[see e.g.,][]{massaro12a,dabrusco13,massaro13c} with different procedures developed and follow up observations 
(see e.g., \'Alvarez Crespo et al. 2016a, \'Alvarez Crespo et al. 2016b).

 

Here we combine the study of the IR colors of the BCUs with the search of the optical spectra available  in the latest releases of Sloan Digital Sky Survey \cite[SDSS DR12,][]{alam15} 
and Six-Degree Field Galaxy Survey Database \cite[6dFGS DR3,][]{jones09}, aiming to confirm their BL Lac or FSRQ nature and estimate their redshifts. 

This updated search in the literature is complementary to our investigation of multifrequency
observations \citep[see e.g,][]{cowperthwaite13,massaro13a,paggi13,massaro13b}
and to the optical spectroscopic campaign carried out to confirm the blazar-like nature of BCUs and
$\gamma$-ray blazar candidates selected to be potential counterparts of the UGSs \citep[see e.g.,][]{paggi14,massaro15,landoni15,ricci15}.
  
 This paper is organized as follows: Section ~\ref{sec:sample} describes the sample considered for the analysis and the underlying selection criteria; 
Section ~\ref{sec:results} is devoted to the archival optical analysis. Section ~\ref{sec:details} details the emission/absorption features of the sources.
Finally, the summary and conclusions are given in Section ~\ref{sec:conclusions}. 

 We use cgs units unless otherwise stated. Spectral indices, $\alpha$, are defined by flux density $S_\nu \propto \nu^{-\alpha}$ and flat spectra when  sources with $\alpha < 0.5$. 
 \wse\ magnitude at [3.4], [4.6], [12], [22] $\mu$m (i.e., the nominal bands) are in the Vega system.
WISE magnitudes are not corrected for the Galactic extinction since, as shown in our previous analyses, such corrections only affects significantly (i.e., but less than $\sim$3\%) 
 the magnitude at 3.4$\mu$ for sources lying at low Galactic latitudes~\citep[see e.g.][]{dabrusco14}. This will also allow us a direct comparison with color-color diagrams published in previous works \citep{massaro12a,massaro12b}.

\section{Sample selection}
\label{sec:sample}
Our initial sample consists of  585  BCUs selected from the 3FGL. 
We first searched for the counterpart in the WISE all-sky survey  in the AllWISE Source Catalog  \cite{cutri13}  with a search radius of 3.3" \cite[see][]{dabrusco14}, 
to test whether their IR colors are consistent with those of  known Fermi blazars \cite{massaro11}.     
In Figures ~\ref{fig:colors_bzb} and ~\ref{fig:colors_bzq} we overlay the IR counterpart of the BCUs in the   [3.4] - [4.6] vs  [4.6] - [12] color-color plane,
locating those compatible with known $\gamma$-ray emitting 
BZBs and BZQs. The BZQs are located in the redder part of the locus occupied by the gamma-ray emitting blazars, while the BL Lacs lie on the bluer part, according to Massaro et al. (2012a).
 Afterwards 
 we crossmatched the sky positions of their radio counterparts with 
SDSS DR12 and 6dFGS DR3 within 2 arcsec radius chosen,
searching for the sources in the footprint of both surveys. We found 124 unique matches in the SDSS and 73 in the 6dFGS for the BCUs selected within 2 arcsec,
however out of these crossmatches only 15 in SDSS and 22 on 6dFGS had a spectrum available

SDSS DR12 represents the culmination of the third phase of the survey. It includes all data taken through 14 July 2014, and encompasses more than 1/3 of the entire celestial sphere, more than four million spectra (see Alam et al. 2015 for more details). 
On the other hand 6dFGS is a combined redshift and peculiar velocity survey which covers most of the southern sky, excluding galactic latitudes $|b|<10$ degrees \cite[see ][]{jones04,jones09}.

We collected spectra and classified each source, determining the redshift in the cases where emission/absorption features were clear. 
We classify the source as noted above: as a BZB if the $EW_{rest} < 5 \AA$ when detected, as a BZG if the emission is dominated by the host elliptical galaxy rather than by non-thermal continuum  arising from the jet \cite{massaro12a};
and when the $EW_{rest} > 5\AA$ and flat radio spectrum, it appears as a BZQ.

In Table 1 we report the 1FGL, 2FGL and 3FGL names together with their classifications and the assigned counterpart for each release of the Fermi catalogs.
In Table 2  we report the name of the BCUs in both 3FGL and WISE all-sky survey in the Allwise Source catalog\footnote{http://wise2.ipac.caltech.edu/docs/release/allwise/}
together with the survey name, the quality of the spectra for 6dFGS or the signal to noise ratio for SDSS, the  classification and the redshift. In Table 3 we report the BCUs already
included in the BZCAT, with the 3FGL name, the WISE all-sky survey name, the telescope, the quality of the spectra for the cases in which it was found in the 6dFGS, the classification,
the redshift and some notes about each source.

\section{Archival Optical Analysis}
\label{sec:results}
We visually inspected all the optical spectra  to avoid misclassifications due to artefacts from the automated spectral  analysis and line identifications. 
The SDSS and 6dFGS classification performed by their automatic procedures sometimes give a value for the redshift for BZBs even though there is no clear evidence of emission/absorption lines. We ignore them, using 
our own values.

We emphasize that in the 6dFGS, redshift measurements are obtained semi-automatically, and spectra are assigned a quality value Q based on visual inspection
 on a scale of 1 to 6 through assessment of every redshift.  Q = 1 is assigned to unusable measurements, Q = 2 to possible but unlikely redshifts, Q = 3 for reliable redshifts and Q = 4 for high-quality redshifts
  \cite{jones09}. The quality of the spectra in the 6dFGS does not stand for the quality in terms of signal to noise as
it is required in the 3LAC for the classification of a source as a BCU I, 
but only a visual assessment of the redshift.  Since by definition most BL Lacs lack of emission/absorption features, their redshifts cannot be measured, so it is expected
that the quality given by 6dFGS is Q=1,2.
To confirm this statement, we searched for sources classified as BL Lacs in the BZCAT in the footprint of 6dFGS, 
and checked the quality of the spectra. From the 82 BL Lacs with a spectrum in the 6dFGS, 51 have Q=1, 16 Q=2,
10 Q=3 and 5 Q=4. Consequently, even if  Q=1 or Q=2, if we do not see any feature we classify the source as BZB.

Nine sources of our initial sample were already reported in the \bzcat, five of them outside the footprint of both sources considered. 
We include them in our analysis because they became available after the 3LAC publication, so we can update the latest release of the Fermi AGN catalog. For those outside the
footprint we found their spectra in the literature and we were able to update the 3LAC classification. We report these \bzcat\ sources on Table (3) and the literature for which we found information is in
the column (7), together with the \bzcat\ name.

We found that 28 sources are BZBs, and confirm/estimate the redshift only for 3 of them. Eight sources are classified as BZQs because they have a flat spectrum ($\alpha < 0.5$)
in the radio band at 1-8 GHz and/or even at lower frequencies \citep{massaro13b,massaro13d}.
The remaining six sources are classified as BZGs. 

\section{Source Details}
\label{sec:details}
Here we report the details of the emission/absorption features found in each source. If the object is a featureless BZB, we do not mention it.

3FGL J0009.6-3211 was classified as a BZG at $z$=0.02 because of the absorption features G band ($\lambda=4416.9\AA, EW=18.1\AA$), Mg ($\lambda=5312.7\AA, EW=4.9\AA$)
and Na ($\lambda=6052.9\AA, EW=3.3\AA$); and the emission line H$\alpha$ ($\lambda=6763.8\AA, EW=1.8\AA$).

3FGL J0028.8+1951 was found to be a BZQ at $z$=1.55 because of the features C IV ($\lambda=3949.9\AA, EW=106.2\AA$); He II ($\lambda=4178.5\AA, EW=4.8\AA$); [C III]  ($\lambda=4862.6\AA, EW=17.0\AA$); 
Mg ($\lambda=7137.9\AA, EW=29.3\AA$); [O III] ($\lambda=9494.6\AA, EW=8.6\AA$).

We found for the source 3FGL J0339.2-1738 the absorption features Ca H\&K ($\lambda=3630.9\AA, EW>50.6\AA$), Mg ($\lambda=3630.9\AA, EW>50.6\AA$), Na ($\lambda=3630.9\AA, EW>50.6\AA$).
It is a BZG at $z$=0.06. 

For the source 3FGL J0904.3+4240 we find the following features: C IV ($\lambda=3630.9\AA, EW>50.6\AA$); [C III] ($\lambda=4462.1\AA, EW=39.9\AA$); Mg ($\lambda=6562.6\AA, EW=61.4\AA$) and 
O II ($\lambda=8727.6\AA, EW=13.9\AA$). Then we classify the source as a BZQ at $z$=1.34 

In 3FGL J1003.6+2608 assuming the line identified is O II ($\lambda=7194.9\AA, EW=4.9\AA$) the redshift estimate is  $z$=0.93 for the BZB.

In 3FGL J1315.4+1130 there is an absorption
feature, CA H\&K ($\lambda = 6801.7-6868.3 \AA$, $EW_{obs} = 2.0 -  4.4 \AA$) that enables us to measure a redshift of $z$ = 0.73. and to classify it as a BZB.

We classify the source 3FGL J1322.1+0838 as a BZQ at $z$=0.32 because of the identification of the lines Mg ($\lambda=3683.4 \AA, EW>48.7\AA$); 
Ca H\&K ($\lambda=5218.6 - 5268.4 \AA, EW=1.6 - 1.1 \AA$); [O III] ($\lambda=6638.5 \AA, EW=1.8\AA$); H$\alpha$ ($\lambda=8726.1 \AA, EW=4.7\AA$).
 
For 3FGL J1342.7+0945 we were able to identify the absorption features Ca H\&K ($\lambda=5048.5 - 5091.6 \AA, EW=1.8 - 2.0 \AA$) and Mg ($\lambda=6636.9 \AA, EW=1.9\AA$);
 and the emission lines O II ($\lambda=4782.3\AA, EW=2.4\AA$) and H$\alpha$ ($\lambda=8436.4\AA, EW=5.9\AA$).
 It is a BZG at $z$ = 0.28.
 
 In 3FGL J1412.0+5249, due to the absorption features Ca H\&K ($\lambda=4236.1 - 4271.9 \AA, EW=8.7 - 6.0 \AA$); G band ($\lambda=4633.3\AA, EW=6.5\AA$);
 Mg ($\lambda=5571.1\AA, EW=6.0\AA$) and Na ($\lambda=6325.9\AA, EW=3.9\AA$),
 we classify it as a BZG at $z$=0.08.
 
The source 3FGL J2346.7+0705 is a BZB, but due to the absorption feature Ca H\&K ($\lambda=4612.2 - 4649.0 \AA, EW=0.3 - 0.9 \AA$) it is possible to estimate a redshift of $z$=0.17.

\begin{figure*}
  \begin{center}
 \includegraphics[width=7.5cm,angle=270]{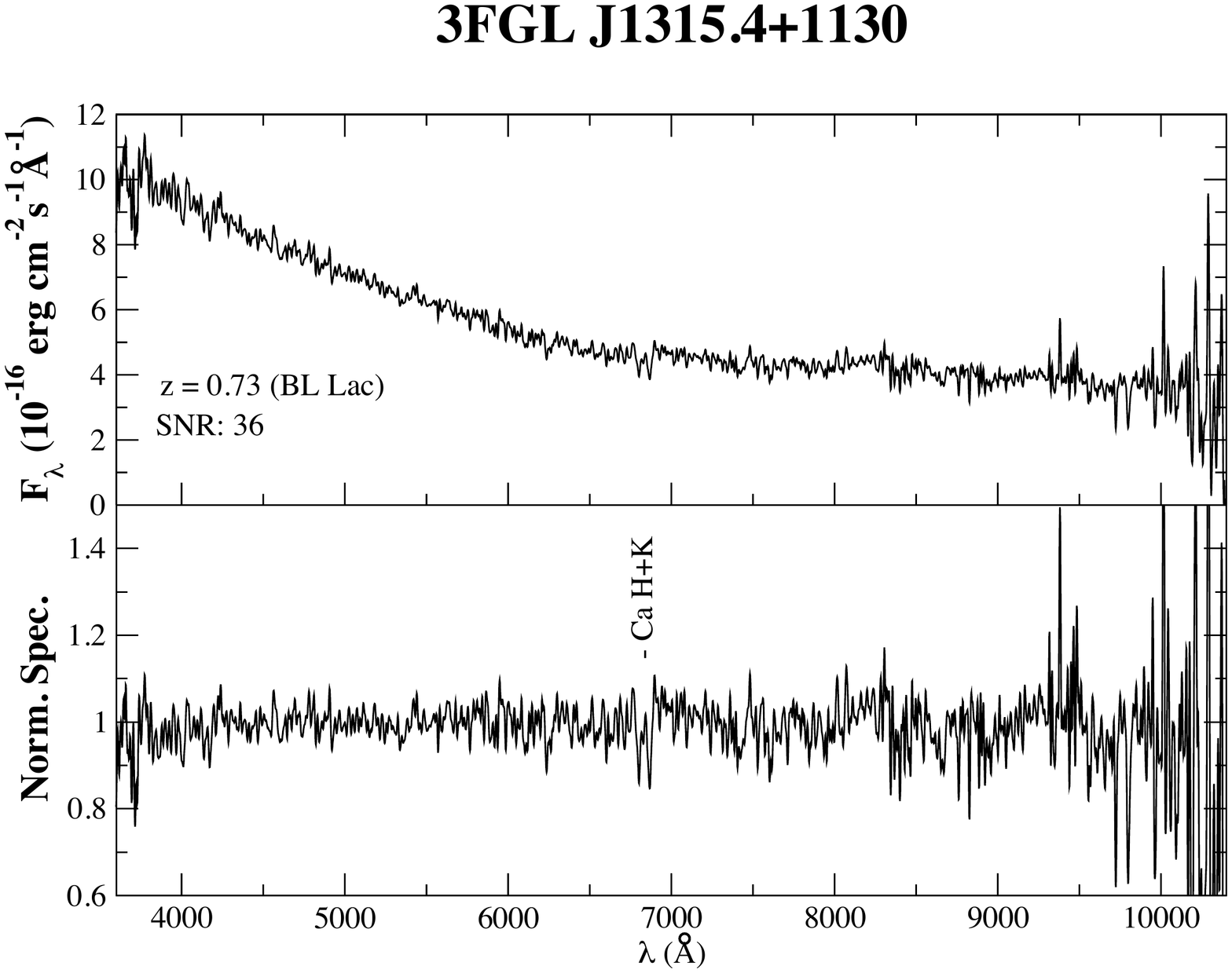}
 \caption{The optical spectrum of WISE J131532.62+113331.7 from the SDSS, potential counterpart of 3FGL J1315.4+1130. It is classified as a BZB at z=0.7 on the basis of its featureless
 optical spectrum, dominated by non-thermal emission from the jet. The average signal to noise is 36.}
 \label{fig:BZB}
 \end{center}
 \end{figure*}

\begin{figure*}
  \begin{center}
 \includegraphics[width=7.5cm,angle=270]{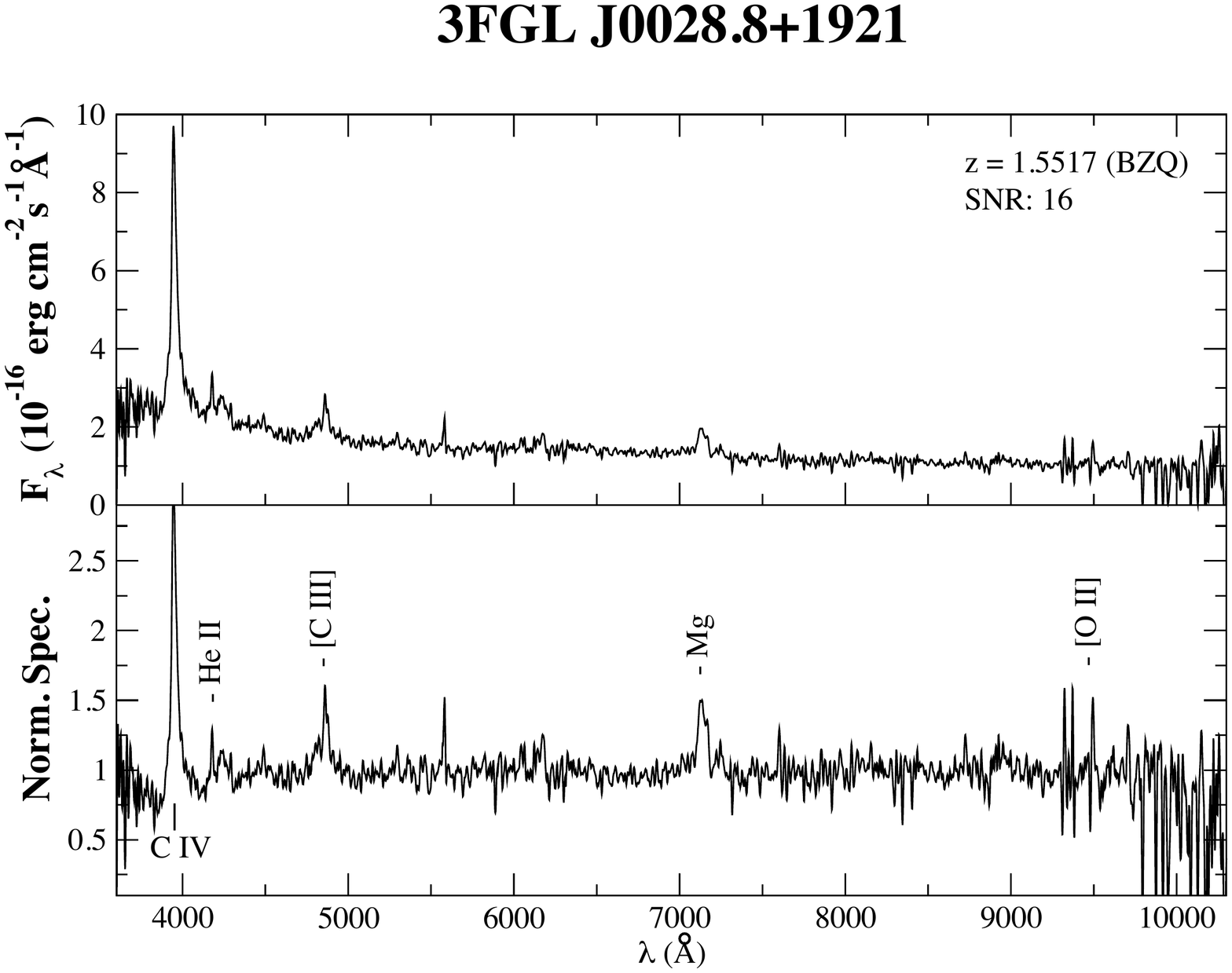}
 \caption{The optical spectrum of WISE J002829.81+200026.7 from the SDSS, potential counterpart of 3FGL J0028.8+1951. Classified as a BZQ at $z$=1.5517 because
 of the identification of the lines C IV, He  II, [C III], Mg and [O II]. The average signal to noise is 16.} 
 \label{fig:BZQ}
 \end{center}
 \end{figure*}

\begin{figure*}
   \begin{center}
 \includegraphics[width=7.5cm,angle=270]{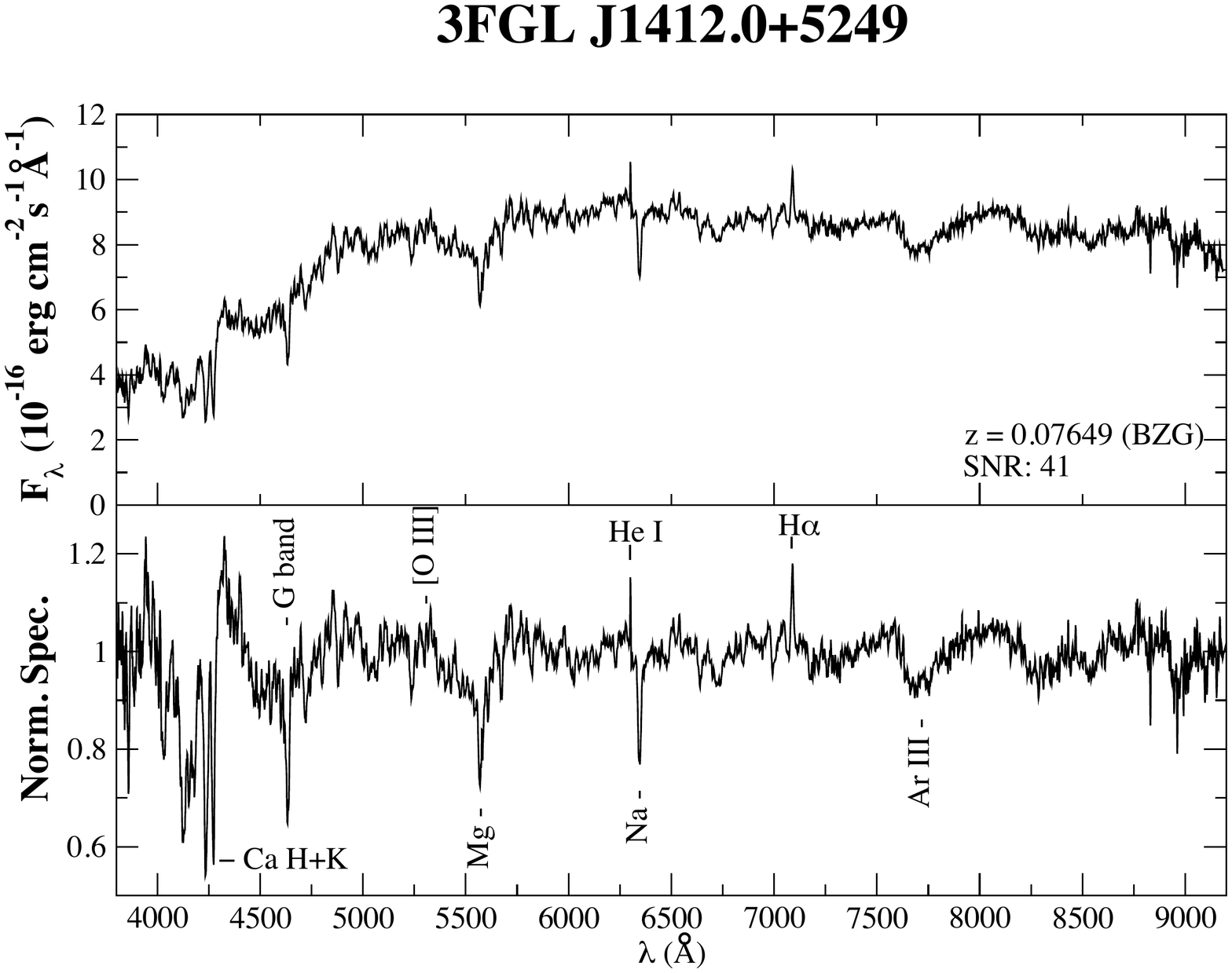}
 \caption{The optical spectrum of WISE J141149.44+524900.2 from the SDSS, potential counterpart of 3FGL J1412.0+5249. The spectrum is dominated by the emission of the host elliptical galaxy and shows the doublet Ca H+K,  the G band, [O III], Mg, He I, Na, H$\alpha$ and Ar III. These features corresponds to a redshift of z = 0.07649. The average signal to noise is 41} 
 \label{fig:BZG}
\end{center}
 \end{figure*}

In Figures ~\ref{fig:BZB}, ~\ref{fig:BZQ} and ~\ref{fig:BZG}  we show the optical spectra for each one of the classifications of our analysis, as an example of a BL Lac, a BZQ and a BZG.

\section{Summary and conclusions}
\label{sec:conclusions}
\begin{figure*}
 \begin{center}
 \includegraphics[height=8.5cm,width=9.5cm]{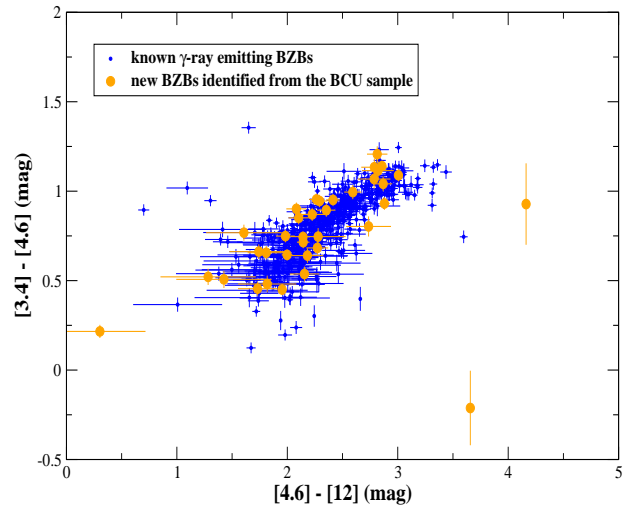}
 \caption{Projection of the BZBs of our sample in the WISE gamma-ray strip in the [3.4] - [4.6] versus [4.6] - [12] color-color plane. } 
 \label{fig:colors_bzb}
 \end{center}
 \end{figure*}

\begin{figure*}
  \begin{center}
 \includegraphics[height=8.5cm,width=9.5cm]{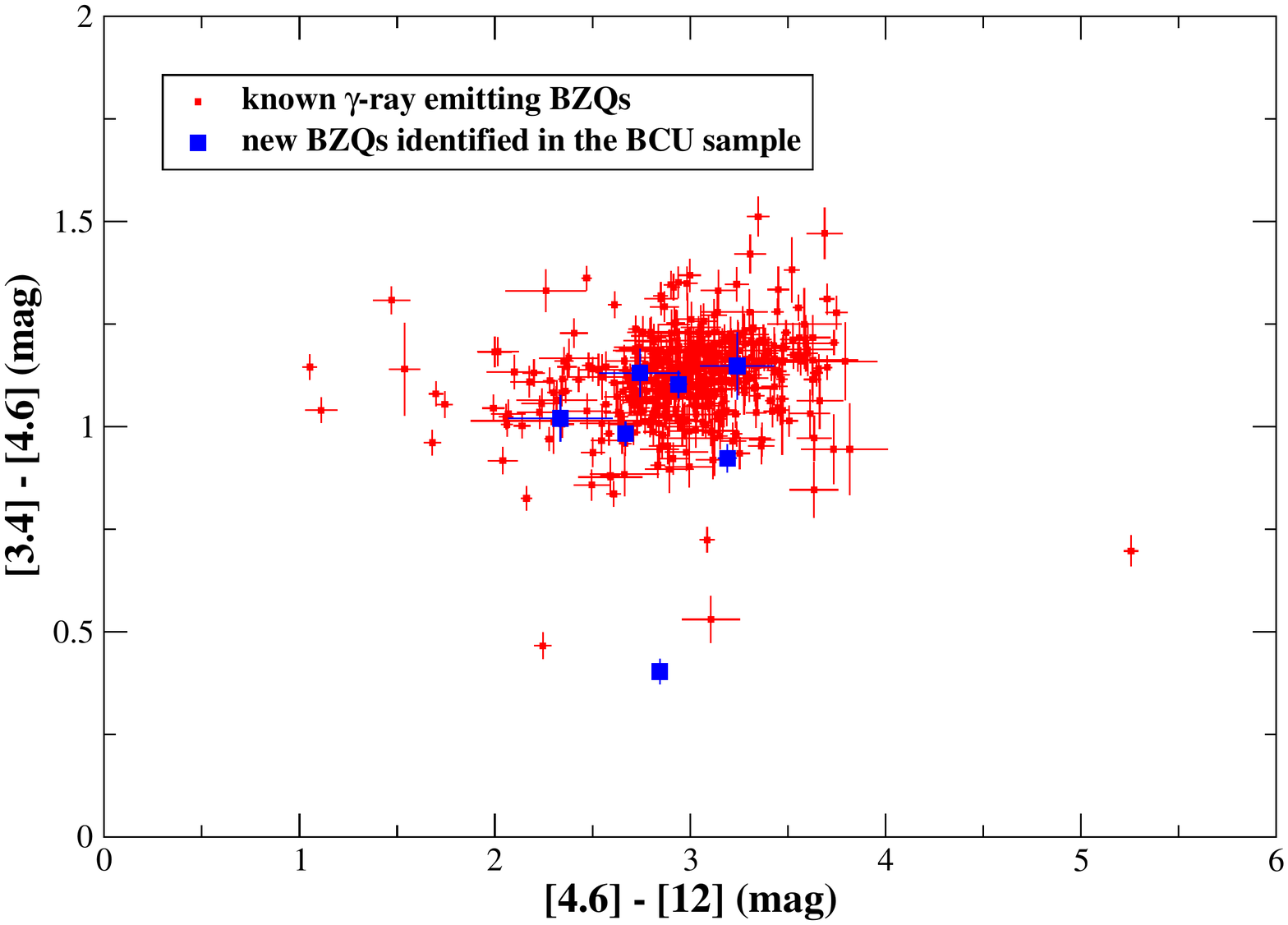}
 \caption{Projection of the BZQs of our sample in the WISE gamma-ray strip in the [3.4] - [4.6] versus [4.6] - [12] color-color plane. } 
 \label{fig:colors_bzq}
 \end{center}
 \end{figure*}

We performed a combined analysis of the mid-IR properties of the 43 BCUs in the 3LAC with the optical spectra.
In Figures ~\ref{fig:colors_bzb} and~\ref{fig:colors_bzq}  we plot the positions of the BZBs and BZQs respectively,  in the WISE IR color-color space in comparison with the $\gamma$-ray blazars from the WIBRaLS catalog 
that define the so called locus \cite[peculiar position of blazars in the IR color-color space,][]{dabrusco14}.
The BZQs are located in the redder part of the locus, and the BL Lacs on the bluer part.

For those sources consistent with being a BL Lac or a FSRQ, 
we searched for archival optical spectra in both surveys SDSS DR12 \cite{alam15} and 6dFGS DR3 \cite{jones09}  to confirm their nature and whenever possible to
estimate their redshifts.

Based on the spectra of both surveys, we classified 28 sources as BL Lacs and determined the redshift for 3 of them, 
8 as FSRQ for which there was radio information available in the literature and 6 BZGs. 
We corrected the automatic classification and redshift measurements for 
3FGL J1315.4+1130. It is classified as a QSO as SDSS, but measuring the $EW_{rest}$ of the emission lines we detected it is indeed a BL Lac, 
and we were able to measure a redshift of $z$ = 0.73 instead of the $z$ = 1.56 given
by their automated classification.

We emphasize that our analysis will be useful not only to reduce the number of unclassified sources
in the future release for the \fer\ catalogs, but to improve our knowledge on the
luminosity function of each of the different blazar classes \citep[see e.g.,][]{ajello12} and thus help determina their
contributions to the extragalactic background \citep[see e.g.,][]{ajello14}. This will translate into more
stringent limits on the dark matter annihilation \citep{ajello15,ackermann15} as well as
on the imprint of the extragalactic background light on BL Lac spectra \citep[see e.g.,][]{ackermann12}.
Our investigation will also be useful to select of potential targets for the Cherenkov Telescope Array (CTA)
\citep{massaro13d,arsioli15} in the near future.


%
%

%
%

%

%

\clearpage
%
This investigation is supported by the NASA grants NNX12AO97G and NNX13AP20G.
The work by G. Tosti is supported by the ASI/INAF contract I/005/12/0.
H. A. Smith acknowledges partial support from NASA/JPL grant RSA 1369566 and NASA grant NNX14.
HOF was funded by a postdoctoral UNAM grant and is currently granted by a C\'atedra CONACyT para J\'ovenes Investigadores.
V. Chavushyan acknowledges funding by CONACyT research grant 151494 (M\'exico).
Part of this work is based on archival data, software or on-line services provided by the ASI Science Data Center.
This research has made use of data obtained from the high-energy Astrophysics Science Archive
Research Center (HEASARC) provided by NASA's Goddard Space Flight Center; 
the SIMBAD database operated at CDS,
Strasbourg, France; the NASA/IPAC Extragalactic Database
(NED) operated by the Jet Propulsion Laboratory, California
Institute of Technology, under contract with the National Aeronautics and Space Administration.
This publication makes use of data products from the Wide-field Infrared Survey Explorer, 
which is a joint project of the University of California, Los Angeles, and 
the Jet Propulsion Laboratory/California Institute of Technology, 
funded by the National Aeronautics and Space Administration.
Funding for the Sloan Digital Sky Survey IV has been provided by the Alfred P. Sloan Foundation, the 
U.S. Department of Energy Office of Science, and the Participating Institutions. SDSS- IV 
acknowledges support and resources from the Center for High-Performance Computing at the University 
of Utah. The SDSS web site is www.sdss.org.
SDSS-IV is managed by the Astrophysical Research Consortium for the Participating Institutions of
the SDSS Collaboration including the Brazilian Participation Group, the Carnegie Institution for 
Science, Carnegie Mellon University, the Chilean Participation Group, the French Participation 
Group, Harvard-Smithsonian Center for Astrophysics, Instituto de Astrof\'isica de Canarias, the Johns 
Hopkins University, Kavli Institute for the Physics and Mathematics of the Universe (IPMU) / 
University of Tokyo, Lawrence Berkeley National Laboratory, Leibniz Institut fur Astrophysik Potsdam 
(AIP), Max-Planck-Institut für Astronomie (MPIA Heidelberg), Max-Planck-Institut fur Astrophysik 
(MPA Garching), Max-Planck-Institut fur Extraterrestrische Physik (MPE), National Astronomical 
Observatory of China, New Mexico State University, New York University, University of Notre Dame, 
Observatorio Nacional / MCTI, the Ohio State University, Pennsylvania State University, Shanghai 
Astronomical Observatory, United Kingdom Participation Group, Universidad Nacional Aut\'onoma de 
M\'exico, University of Arizona, University of Colorado Boulder, University of Oxford, University of 
Portsmouth, University of Utah, University of Virginia, University of Washington, University of 
Wisconsin, Vanderbilt University, and Yale University.
TOPCAT\footnote{\underline{http://www.star.bris.ac.uk/$\sim$mbt/topcat/}} 
\citep{taylor05} for the preparation and manipulation of the tabular data and the images.


%


%
%

\clearpage

\begin{table*}
 \begin{center}
\tiny
 \caption{Gamma-ray Name, Classification and Counterparts in the  \fer\ Catalogs}
\label{tab:log1}
\begin{tabular}{|lllllllll|}
\hline
1FGL & 1FGL & 1FGL  & 2FGL & 2FGL & 2FGL  & 3FGL & 3LAC & 3FGL \\
name & class & ctp      & name & class & ctp            & name & class & ctp \\
\hline
\hline 
   &   &   &   &   &     & J0003.8-1151 & bcu II & PMN J0004-1148 \\
J0008.9+0635 & bll & CRATES J0009+0628 & J0009.0+0632 & bll & CRATES J0009+0628    & J0009.6-3211 & bcu I & IC 1531  \\
  &   &   &   &   &   &    J0028.8+1951 & bcu III & TXS 0025+197 \\
    &   &   &   &   &    & J0030.2-1646 & bcu II & 1RXS J003019.6-164723  \\
    &   &   &   &    &   & J0043.5-0444 & bcu II & 1RXS J004333.7-044257  \\
    &   &   & J0146.6-5206 &  agu & PKS 0144-522 &   J0147.0-5204 & bcu I & PKS 0144-522  \\
        & &   & & &  & J0156.9-4742 & bcu II & 2MASS J01564603-4744174 \\
    & &   & &   & &    J0255.8+0532 & bcu II & PMN J0255+0533 \\
    & &   & &   & &    J0301.8-7157 & bcu II & PKS 0301-721 \\
  J0339.1-1734 & agn & PKS0336-177 & J0339.2-1734 & agn & PKS 0336-177 &  J0339.2-1738 & bcu I & PKS 0336-177 \\
    & &   & &   & &   J0343.3+3622 & bcu I & OE 367 \\
    & &   & J0433.4-6029 & agu & PKS 0432-606 &  J0433.7-6028 & bcu II & PKS 0432-606 \\
     & &   & J0435.1-2341 & agu & PMN J0434-2342 &  J0434.4-2341 & bcu I & PMN J0434-2342 \\
  J0439.8-1857 &   & & J0439.8-1858 &   &  & J0439.9-1859 & bcu II & PMN J0439-1900 \\
  J0521.6+0103 &   & & J0521.9+0108 & fsrq & PKS0519+01 &  J0521.7+0103 & bcu II & NVSS J052140+010257 \\
    & &   & J0730.6-6607 &  agu & PMN J0730-6602  & J0730.5-6606 & bcu II & PMN J0730-6602 \\
    & &   & &   &    & J0827.2-0711 & bcu I & PMN J0827-0708 \\
    & &   & J0903.6+4238 &  agn & S4 0900+42 &  J0904.3+4240 & bcu II & S4 0900+42 \\
    & &   & &   & &    J0921.0-2258 & bcu II & NVSS J092057-225721 \\
    & &   & &   & &    J1003.6+2608 & bcu I& PKS 1000+26 \\
  J1040.5+0616 &   & & J1040.7+0614 & fsrq & 4C +06.41 &  J1040.4+0615 & bcu II & GB6 J1040+0617 \\
    & &   & &   & &   J1040.8+1342 & bcu II & 1RXS J104057.7+134216 \\
    & &   & &   & &   J1125.0-2101 & bcu II & PMN J1125-2100 \\
    & &   & &   & &   J1129.4-4215 & bcu I & SUMSS J113006-421441 \\
    & &   & &   & &   J1200.8+1228 & bcu II & GB6 J1200+1230 \\
  J1256.5-1148 & agu & CRATES J1256-1146 & J1256.5-1145 & agn & PMN J1256-1146  & J1256.3-1146 & bcu I & PMN J1256-1146 \\
    & &   & &   & &    J1315.4+1130 & bcu II & 1RXS J131531.9+113327 \\
  J1322.1+0838 &   & &   & &   &  J1322.3+0839 & bcu II & NVSS J132210+084231 \\
  J1340.5-0413 &   & & J1340.5-0412 &   & &    J1340.6-0408 & bcu II & NVSS J134042-041006 \\
    & &   & &   &    & J1342.7+0945 & bcu II & NVSS J134240+094752 \\
    & &   & &   & &    J1412.0+5249 & bcu I & SBS 1410+530 \\
    & &   & &   & &   J1549.5+1709 & bcu II & MG1 J154930+1708 \\
    & &   & &   & &   J1636.7+2624 & bcu II & NVSS J163651+262657 \\
    & &   & J1656.9-2008 & agu & 1RXS J165655.0-201049 &  J1656.8-2010 & bcu II & 1RXS J165655.0-201049 \\
    & &   & J1954.4-1607 &  agu & 1RXS J195500.6-160328  & J1955.0-1605 & bcu II & 1RXS J195500.6-160328 \\
    & &   & &   & &    J2104.2-0211 & bcu II & NVSS J210421-021239 \\
  J2108.6-6646 & agu & PKS 2104-668 & J2108.9-6636 & agu & PKS 2104-668 & J2109.1-6638 & bcu II & PKS 2104-668 \\
  J2118.3-3237 &   & &   & &   &  J2118.0-3241 & bcu I & NVSS J211754-324326 \\
     & &   & &   & &   J2232.9-2021 & bcu II & 1RXS J223249.5-202232 \\
    & &   & J2317.3-4534 & agu & 1RXS J231733.0-453348 &  J2317.3-4534 & bcu II & 1RXS J231733.0-453348 \\
    & &   & J2327.9-4037 & agu & PKS 2325-408 &  J2328.4-4034 & bcu II & PKS 2325-408 \\
    & &   & &   & & J2346.7+0705 & bcu II & TXS 2344+068 \\
\noalign{\smallskip}
\hline
\hline 
\end{tabular}\\
bcu = AGN of uncertain type;  bzb = bll = BL Lac; bzq = blazar of QSO type
Column description. (1): 1FGL name, (2): 1FGL classification, (3): 1FGL assigned counterpart, (4): 2FGL name, 
(5): 2FGL classification, (6): 2FGL assigned counterpart,  (7): 3FGL name,
(8): 3LAC classification, (9): 3FGL assigned counterpart.
\end{center} 
\end{table*}

\begin{table*}
 \begin{center}
\tiny
 \caption{}
\label{tab:log1}
\begin{tabular}{|llccccl|}
\hline
3FGL              & WISE          	    & Survey$\backslash$telescope  	& Q & SN  & classification     & $z$\\
name               & name     		    & name						& 6dFG & SSDS   & 	          &       \\
\hline
\hline 
  J0009.6-3211 & J000935.55-321636.8 & 6dFGS & 4  & - & bzg & 0.02\\
  J0028.8+1951 & J002829.81+200026.7 & SDSS & - & 24 & bzq & 1.55\\
  J0030.2-1646 & J003019.40-164711.7 & 6dFGS & 1 & - & bzb &  \\
  J0043.5-0444 & J004334.12-044300.6 & 6dFGS & 1 & - & bzb & 1.63  \\
  J0156.9-4742 & J015646.03-474417.3 & 6dFGS & 1 &  - & bzb &  \\
  J0255.8+0532 & J025549.51+053355.0 & SDSS & - & 32 & bzb &   \\
  J0339.2-1738 & J033913.70-173600.8 & 6dFGS & 4 & - & bzg & 0.06 \\
  J0439.9-1859 & J043949.72-190101.5 & 6dFGS & 2 & - & bzb &  \\
  J0521.7+0103 & J052140.82+010255.5 & SDSS & - & 27 & bzb & \\
  J0730.5-6606 & J073049.51-660218.9 & 6dFGS & 4 &- & bzb &  \\
  J0827.2-0711 & J082706.16-070845.9 & 6dFGS & 1 & - & bzb & \\
  J0904.3+4240 & J090415.62+423804.5 & SDSS &  - & 15 & bzq & 1.34 \\
  J0921.0-2258 & J092057.47-225721.5 & 6dFGS & 1 & - & bzb & \\
  J1003.6+2608 & J100342.22+260512.8 & SDSS & -& 8 & bzb & 0.93\\
  J1040.4+0615 & J104031.62+061721.7 & SDSS & - & 9 & bzb & \\
  J1040.8+1342 & J104057.69+134211.7 & SDSS & - & 16 & bzb &  \\
  J1125.0-2101 & J112508.62-210105.9 & 6dFGS & 1 & - & bzb &   \\
  J1129.4-4215 & J113007.04-421440.9 & 6dFGS & 3 & - & bzb &  \\
  J1200.8+1228 & J120040.03+123103.2 & SDSS & - & 16 & bzb &  \\
  J1315.4+1130 & J131532.62+113331.7 & SDSS & - & 13 & bzb & 0.73  \\
  J1322.3+0839 & J132210.17+084232.9 & SDSS & - & 41 & bzq & 0.32 \\
  J1340.6-0408 & J134042.02-041006.8 & 6dFGS & 2 & - & bzb & \\
  J1342.7+0945 & J134240.02+094752.4 & SDSS & - & 32 & bzg & 0.28  \\ 
  J1412.0+5249 & J141149.44+524900.2 & SDSS & - & 44 & bzg & 0.08 \\
  J1549.5+1709 & J154929.28+170828.1 & SDSS & - & 6 & bzb & \\
  J1636.7+2624 & J163651.46+262656.7 & SDSS & - & 21 & bzb & \\
  J1656.8-2010 & J165655.14-201056.2 & 6dFGS & 2 & - & bzb & \\
  J1955.0-1605 & J195500.65-160338.4 & 6dFGS & 1 & - & bzb & \\
  J2104.2-0211 & J210421.92-021239.0 & 6dFGS & 2 & - & bzb & \\
  J2118.0-3241 & J211754.91-324328.2 & 6dFGS & 4 & - & bzb & \\
  J2232.9-2021 & J223248.80-202226.2 & 6dFGS & 1 & - & bzb & \\ 
  J2317.3-4534 & J231731.98-453359.6 & 6dFG & 1 & - & bzb & \\
  J2346.7+0705 & J234639.93+070506.8 & SDSS & - & 73 & bzb & 0.17 \\
\noalign{\smallskip}
\hline
\hline 
\end{tabular}\\
Column description: (1): 3FGL name ($^*$) indicates sources in the in the  Roma-BZCAT, (2): WISE name, (3): Survey/telescope: Six-Degree Field Galaxy Survey Database (6dFG), Sloan Digital Sky Survey (SDSS), (4): quality of the spectra in 6dFGS, (5): signal to noise ratio in the SDSS, (6): source classification, (7): redshift.
\end{center}
\end{table*}

\begin{table*}
 \begin{center}
\tiny
 \caption{}
\label{tab:log2}
\begin{tabular}{|llcccll|}
\hline
3FGL              & WISE          	    & Survey$\backslash$telescope  	& Q & classification     & $z$ & Notes\\
name               & name     		    & name						& 6dFGS  	          & 	& \\
\hline
\hline 
  J0003.8-1151 & J000404.91-114858.3  & 6dFGS & 4 &  bzb &   & 5BZB J0004-1148 \\
  J0147.0-5204 & J014648.58-520233.5 & 6dFGS & 4 & bzg & 0.09  & 5BZG J0146-5202 \\
  J0301.8-7157$^*$ & J030138.47-715634.5 & GemN. & & bzq & 0.82 & Titov+13, 5BZQ J0301-7156\\
  J0343.3+3622$^*$ & J034328.94+362212.4 & Palomar  & & bzq & 1.48 & Vermeulen+95, 5BZQ J0343+3622\\	
  J0433.7-6028$^*$ & J043334.59-603010.3 & GemN.  & & bzq & 0.93 & 5BZQ J0433-6030 \\ 
  J0434.4-2341$^*$ & J043428.98-234205.3 & NTT & & bzb &   & Shaw+13, 5BZB J0434-2342 \\
  J1256.3-1146 & J125615.95-114637.3 & 6dFGS & 4 & bzg & 0.06  & 5BZG J1256-1146\\
  J2109.1-6638$^*$ & J210851.80-663722.7 & 6dFGS+GemN. & & bzb &  & Titov+13, 5BZB J2108-6637\\
  J2328.4-4034$^*$ & J232819.26-403509.8 & GemN. & &  bzq & 1.97 & 5BZQ J2328-4035 \\ 
\noalign{\smallskip}
\hline
\hline 
\end{tabular}\\
Column description: (1): 3FGL name ($^*$ indicates sources in the in the  Roma-BZCAT), (2): WISE name, (3): Survey/telescope: Six-Degree Field Galaxy Survey Database (6dFGS),  Gemini North telescope (GemN), Palomar Observatory (Palomar), New Technology Telescope (NTT), Multiple Mirror Telescope (MMT), (4): quality of the spectra in 6dFGS, (5): signal to noise ratio in the telesccopes, 
(6): source classification, (7): redshift, (8): notes.
\end{center}
\end{table*}

\end{document}